\documentclass[11pt]{article}

\usepackage[margin=1in]{geometry}
\usepackage{amsmath}
\usepackage{amssymb}
\usepackage{hyperref}
\usepackage{booktabs}
\usepackage{listings}
\usepackage{xcolor}
\usepackage{graphicx}
\usepackage{authblk}
\usepackage{pgfplots}
\usepackage{tikz}
\pgfplotsset{compat=1.18}

\hypersetup{
    colorlinks=true,
    linkcolor=blue,
    citecolor=blue,
    urlcolor=blue
}

\title{GoldbachGPU: An Open Source GPU-Accelerated Framework for\\
       Verification of Goldbach's Conjecture}

\author[1]{Isaac Llorente-Saguer}
\affil[1]{Independent Researcher, London, United Kingdom}
\affil[ ]{\textit{illorentes@gmail.com}}

\date{\today}

\begin{document}
\maketitle

\begin{abstract}
We present \textit{GoldbachGPU}, an open-source framework for large-scale computational verification of Goldbach's conjecture using commodity GPU hardware. Prior GPU-based approaches reported a hard memory ceiling near \(10^{11}\) due to monolithic prime-table allocation. We show that this limitation is architectural rather than fundamental: a dense bit-packed prime representation provides a \(16\times\) reduction in memory footprint, and a segmented double-sieve design removes the VRAM ceiling entirely. By inverting the verification loop and combining a GPU fast-path with a multi-phase primality oracle, the framework achieves exhaustive verification up to \(10^{12}\) on a single NVIDIA RTX~3070 (8\,GB VRAM), with no counterexamples found. Each segment requires \(14\)\,MB of VRAM, yielding \(O(N)\) wall-clock time and \(O(1)\) memory in \(N\). A rigorous CPU fallback guarantees mathematical completeness, though it was never invoked in practice. An arbitrary-precision checker using GMP and OpenMP extends single-number verification to \(10^{10000}\) via a synchronised batch-search strategy. The segmented architecture also exhibits clean multi-GPU scaling on data-centre hardware (tested on 8$\times$H100). All code is open-source, documented, and reproducible on both commodity and high-end hardware.
\end{abstract}

\section{Introduction}
\label{sec:intro}

Goldbach's conjecture, first stated in a 1742 letter from Christian Goldbach to Leonhard Euler, asserts that every even integer greater than~2 can be expressed as the sum of two prime numbers. Despite its elementary formulation, the conjecture remains unproven. Substantial theoretical progress has been made—most notably Chen's theorem~\cite{chen1984}, which shows that every sufficiently large even integer is the sum of a prime and a semiprime, and Helfgott's proof of the ternary Goldbach conjecture~\cite{helfgott2014ternary}—but the binary form continues to resist proof.

Computational verification has therefore played an important complementary role. Over more than a century, increasingly sophisticated algorithms and hardware have pushed the verified range from $10^{4}$ in the nineteenth century to $4\times 10^{18}$ in the state-of-the-art distributed computation of Oliveira~e~Silva, Herzog, and Pardi~\cite{oliveira2014}. Their work relies on highly optimised cache-efficient segmented sieves and large-scale parallelism across many machines over several years.

In contrast, verification on commodity hardware remains limited. The advent of General-Purpose GPU (GPGPU) computing offers substantial parallel throughput, but naive GPU implementations encounter a hard memory wall: storing a monolithic prime table in VRAM becomes infeasible beyond $10^{10}$--$10^{11}$. Prior GPU-based studies~\cite{varon2020,cuda_dp} explicitly identify VRAM exhaustion and thread divergence as the primary obstacles to scaling, and no existing open-source framework demonstrates exhaustive verification beyond $10^{11}$ on a single GPU.

\sloppy
This work addresses that gap. We present \texttt{GoldbachGPU}, an open-source framework that removes the VRAM ceiling through a segmented double-sieve architecture inspired by Oliveira~e~Silva's loop inversion strategy. By combining a dense bit-packed prime representation with a correct segmentation scheme and a multi-phase primality oracle, our implementation achieves exhaustive verification up to $10^{12}$ on a single consumer GPU (NVIDIA RTX~3070, 8\,GB VRAM). A rigorous CPU fallback guarantees mathematical completeness, and an arbitrary-precision GMP-based checker extends single-number verification to $10^{10000}$.

This work makes the following contributions:
\begin{itemize}
    \item A \textbf{segmented double-sieve architecture} that eliminates the VRAM ceiling of prior GPU approaches and enables exhaustive verification to $10^{12}$ on commodity hardware.
    \item A \textbf{dense prime bitset} providing a $16\times$ memory reduction over byte-array representations, enabling GPU-resident prime tables up to $10^{11}$.
    \item A \textbf{two-phase verification design} combining a GPU fast-path with a rigorous CPU exhaustive fallback, ensuring correctness regardless of the growth of the minimal Goldbach prime.
    \item An \textbf{arbitrary-precision checker} using GMP and OpenMP with a synchronised batch-search strategy, verified up to $10^{10000}$.
    \item A fully open-source, documented, and reproducible implementation targeting consumer hardware.
\end{itemize}

\section{Methods}
\label{sec:methods}

The framework consists of five complementary verification tools, arranged in increasing order of capability. Each tool validates against the previous one, establishing a correctness chain from a CPU oracle to arbitrary-precision verification.

\subsection{CPU Baseline (\texttt{cpu\_goldbach})}

A sequential CPU implementation serves as the ground-truth oracle. A segmented Sieve of Eratosthenes generates all primes up to $N$. For each even integer $n \in [4,N]$, the verifier scans primes $p$ from $2$ to $n/2$ and checks whether $q = n - p$ is prime via direct bitset lookup. This establishes a reference implementation against which all GPU tools are validated.

\subsection{GPU Global Bitset Verifier (\texttt{goldbach\_gpu2})}
\label{sec:gpu2}

This tool introduces the dense bitset representation that makes large GPU-resident prime tables feasible.

\paragraph{Dense prime bitset.}
Even integers (except $2$) are never prime and are omitted from storage. For an odd integer $n \ge 3$, the bit index is
\[
\text{word} = \left\lfloor \frac{(n-3)/2}{64} \right\rfloor,
\qquad
\text{bit} = \left( \frac{n-3}{2} \right) \bmod 64,
\]
stored in a 64-bit word array. This achieves a $16\times$ memory reduction relative to byte-per-number storage. At $N = 10^{11}$, the bitset requires $5.96$\,GB of VRAM (within the 8\,GB limit of an RTX~3070), compared to $95$\,GB for a byte array.

\paragraph{Hard VRAM ceiling.}
Because the entire prime table must reside in VRAM, this design cannot exceed $N \approx 10^{11}$ on an 8\,GB GPU. Extending to $10^{12}$ would require $\approx 59$\,GB of VRAM.

\paragraph{GPU kernel.}
One CUDA thread is assigned per even integer $n$. Each thread scans primes $p$ from $2$ to $\min(n/2, N)$ using the bitset. Bounds checking prevents out-of-range memory accesses, and early exit on the first valid partition reduces unnecessary work. The CPU bitset is constructed using a parallelised segmented sieve with OpenMP (20 threads, $1.7\times$ speedup, memory-bandwidth limited).

\subsection{Segmented Double-Sieve Verifier (\texttt{goldbach\_gpu3})}
\label{sec:gpu3}

\texttt{goldbach\_gpu3} is the primary contribution of this work. It removes the VRAM ceiling of\\ \texttt{goldbach\_gpu2} through a correct segmentation strategy and a multi-phase primality oracle.

\subsubsection{Correct segmentation}

A naive segmentation of the prime table is insufficient: for a given even $n$, a valid partition 
$(p,q)$ with $p+q=n$ may place $q$ anywhere on the number line. Any design that forces $q$ to lie 
in the same segment as $p$ will miss valid partitions and therefore cannot be used for exhaustive 
verification.

Following the double-sieve method of Oliveira~e~Silva~\cite{oliveira2014}, we invert the verification 
loop. Instead of iterating over even integers $n$ and searching for a suitable $p$, we iterate over 
candidate primes $p$ and mark all even $n$ for which $q=n-p$ is prime. This inversion allows the 
prime table to be segmented safely. For each segment $[A,B]$, the CPU performs a segmented Sieve 
of Eratosthenes to determine the primality of all odd integers in that segment. The resulting 
bitset---of size $14$\,MB for our fixed segment size 
$\texttt{SEG\_SIZE} = 10^{7}$ even integers---is transferred to VRAM, where it remains 
resident for the duration of the segment's verification.

For each even $n\in[A,B]$, the GPU kernel iterates over candidate primes $p \le P_{\text{SMALL}}$ 
and computes $q=n-p$. The primality of $q$ is resolved by a three-way oracle, evaluated in strict 
priority order:
\begin{enumerate}
    \item $q \le P_{\text{SMALL}}$: lookup in the permanently resident small-primes bitset 
          ($\approx 122$\,KB).
    \item $q \in [A,B]$: lookup in the current segment bitset ($\approx 14$\,MB).
    \item $P_{\text{SMALL}} < q < A$: deterministic 64-bit Miller--Rabin test~\cite{sinclair2011}.
\end{enumerate}

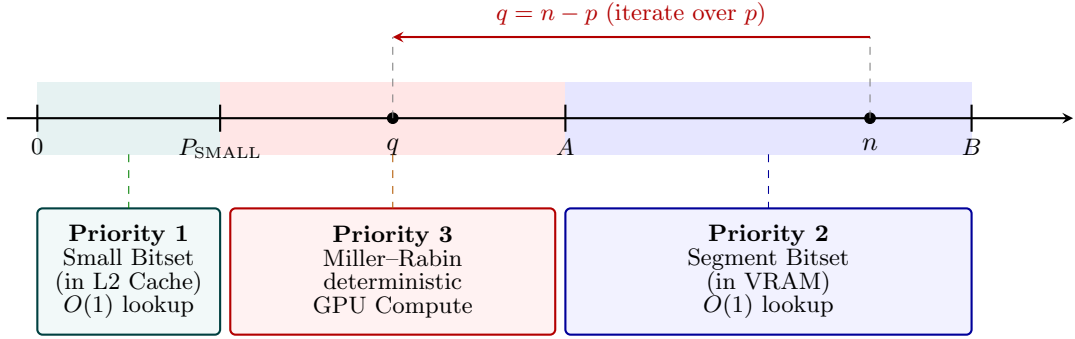
\begin{figure}[ht]
\centering
\begin{tikzpicture}[x=1.35cm, y=1.2cm, font=\sffamily\small, >=stealth]

\def\psmall{1.8}
\def\segA{5.2}
\def\segB{9.2}
\def\npos{8.2}
\def\qpos{3.5}

\fill[teal!10]  (0, -0.4) rectangle (\psmall, 0.4);
\fill[red!10] (\psmall, -0.4) rectangle (\segA, 0.4);
\fill[blue!10]   (\segA, -0.4) rectangle (\segB, 0.4);

\draw[thick, ->] (-0.3, 0) -- (10.2, 0);

\foreach \x/\lbl in {0/0, \psmall/P_{\text{SMALL}}, \segA/A, \segB/B} {
    \draw[thick] (\x, -0.15) -- (\x, 0.15);
    \node[below=4pt, font=\footnotesize] at (\x, 0) {$\lbl$};
}

\filldraw[black] (\npos, 0) circle (2pt) node[below=4pt] {$n$};
\filldraw[black] (\qpos, 0) circle (2pt) node[below=4pt] {$q$};

\draw[gray, dashed, thin] (\npos, 0) -- (\npos, 0.9);
\draw[gray, dashed, thin] (\qpos, 0) -- (\qpos, 0.9);
\draw[->, thick, red!70!black] (\npos, 0.9) -- (\qpos, 0.9) 
    node[midway, above, font=\footnotesize] {$q = n - p$ (iterate over $p$)};

\def\boxy{-1.0}   
\def\boxh{-1.4}   

\draw[teal!50!black, thick, fill=teal!5, rounded corners=2pt] 
    (0, \boxy) rectangle (\psmall, \boxy+\boxh);
\node[align=center, font=\footnotesize] at (\psmall/2, \boxy+0.5*\boxh) {
    \textbf{Priority 1}\\[-2pt]
    Small Bitset\\[-2pt]
    (in L2 Cache)\\[-2pt]
    $O(1)$ lookup
};

\draw[red!70!black, thick, fill=red!5, rounded corners=2pt] 
    (\psmall+0.1, \boxy) rectangle (\segA-0.1, \boxy+\boxh);
\node[align=center, font=\footnotesize] at ({(\psmall+\segA)/2}, \boxy+0.5*\boxh) {
    \textbf{Priority 3}\\[-2pt]
    Miller--Rabin\\[-2pt]
    deterministic\\[-2pt]
    GPU Compute
};

\draw[blue!60!black, thick, fill=blue!5, rounded corners=2pt] 
    (\segA, \boxy) rectangle (\segB, \boxy+\boxh);
\node[align=center, font=\footnotesize] at ({(\segA+\segB)/2}, \boxy+0.5*\boxh) {
    \textbf{Priority 2}\\[-2pt]
    Segment Bitset\\[-2pt]
    (in VRAM)\\[-2pt]
    $O(1)$ lookup
};

\draw[dashed, green!50!black, thin] (\psmall/2, -0.4) -- (\psmall/2, \boxy);
\draw[dashed, orange!70!black, thin] ({(\psmall+\segA)/2}, -0.4) -- ({(\psmall+\segA)/2}, \boxy);
\draw[dashed, blue!60!black, thin] ({(\segA+\segB)/2}, -0.4) -- ({(\segA+\segB)/2}, \boxy);

\end{tikzpicture}
\caption{Primality-oracle selection in \texttt{goldbach\_gpu3}. For each even integer $n$ in the current segment $[A,B]$, the verifier iterates over candidate primes $p$ and computes $q = n - p$. The primality of $q$ is resolved by the first applicable oracle in a strict priority chain designed to minimize GPU compute cost: \textbf{(1)} the resident 122\,KB small-primes bitset when $q \le P_{\text{SMALL}}$ (where $P_{\text{SMALL}}$ is the upper bound of the permanently resident small-primes table); \textbf{(2)} the 14\,MB segment bitset, pre-sieved on the CPU, when $q \in [A,B]$; or \textbf{(3)} a deterministic 64-bit Miller--Rabin test for the transitional region $P_{\text{SMALL}} < q < A$. This three-way structure guarantees complete coverage of all possible $q$ values without requiring a monolithic prime table to reside in VRAM.}
\label{fig:lookup_logic}
\end{figure}

This priority chain covers all possible locations of $q$ without requiring a monolithic prime table 
in VRAM. Because the segment size is fixed, VRAM usage remains constant across all ranges, and 
the lookup logic guarantees correctness regardless of the distribution of Goldbach partitions. 
Figure~\ref{fig:lookup_logic} illustrates the structure of the oracle.

\subsubsection{Phase 1: GPU fast-path and the H(x) envelope}

The worst-case minimal Goldbach prime $H(x) = \max_{n \le x} p_{\min}(n)$ grows extremely slowly. A quadratic fit ($R^2 = 0.991$) to the upper-envelope data of Oliveira~e~Silva~\cite{oliveira2014} gives
\[
H(x) \approx 38.572(\log_{10} x)^2 - 193.21\log_{10} x + 297.36,
\label{eq:hx}
\]
predicting $H(10^{12}) \approx 2000$. We set $P_{\text{SMALL}} = 10^6$, exceeding this bound by multiple orders of magnitude. An atomic counter tracks unresolved $n$ values; after each prime batch, a single 8-byte transfer determines whether further batches are required.

\subsubsection{Phase 2: exhaustive CPU fallback}

If Phase~1 fails to find a partition for some $n$, an exhaustive CPU fallback tests all odd $p$ from $3$ to $n/2$ by trial division, with no upper bound on $p$. This guarantees mathematical completeness regardless of the accuracy of Equation~\ref{eq:hx}. In all experiments reported here, Phase~2 was never invoked.

\subsubsection{Complexity and memory}

Each segment of $S$ even integers is processed in $O(S \cdot \pi(P_{\text{SMALL}}))$ operations, where $\pi(P_{\text{SMALL}})$ is the number of primes up to $P_{\text{SMALL}}$. With $S$ and $P_{\text{SMALL}}$ fixed, the per-segment cost is constant in $N$. Total wall-clock time is therefore $O(N)$, and VRAM usage is $O(1)$ in $N$ (fixed at $14$\,MB per segment). This contrasts with \texttt{goldbach\_gpu2}, which requires $O(N/16)$ bytes of VRAM and cannot exceed $\approx 10^{11}$ on an 8\,GB device.

\subsection{GPU Single-Number Checker (\texttt{single\_check})}

For individual 64-bit even integers (up to $\approx 1.8\times 10^{19}$), \texttt{single\_check} uses a GPU-based deterministic Miller--Rabin test. Small primes $p \le n/2$ are generated on the CPU in batches; for each batch, a grid of GPU threads tests whether $q = n - p$ is prime using the 12-witness deterministic variant proven correct for all 64-bit integers~\cite{sinclair2011}. Due to concurrency, the returned partition is valid but not necessarily minimal.

\subsection{Arbitrary Precision Checker (\texttt{big\_check})}

To verify even integers beyond the 64-bit range, \texttt{big\_check} uses the GNU Multiple Precision library (GMP)~\cite{gmp}. For a given $n$ supplied as a decimal string, small primes $p$ are generated up to $10^7$. For each $p$, the complement $q = n - p$ is computed exactly using GMP arithmetic, and primality of $q$ is tested using GMP's Miller--Rabin implementation~\cite{rabin1980probabilistic} with 25 rounds (false-positive probability $< 10^{-15}$).

\paragraph{Synchronized batch search.}
Naïve OpenMP parallelism causes \emph{thread runaway}: fast threads advance to large primes while slow threads stall on expensive primality tests. To prevent this, primes are processed in batches of 1000 with a synchronization barrier between batches. An atomic flag signals early termination once any thread finds a valid partition. Within a batch, threads exit immediately upon flag detection. This prevents both inter-batch and intra-batch runaway and ensures good load balance via \texttt{schedule(dynamic,1)}.

The practical limit of \texttt{big\_check} is governed by the $O(d^3)$ cost of GMP primality tests in the digit count $d$. At $d = 10{,}001$ ($10^{10000}$), each test takes milliseconds; at $d = 100{,}001$ ($10^{100000}$), each test takes seconds, making exhaustive search infeasible.

\section{Results}
\label{sec:results}

All experiments were conducted on an Intel i7--12700H CPU (20 logical threads, WSL2 Ubuntu~24.04), an NVIDIA RTX~3070 GPU (8\,GB VRAM, 448\,GB/s bandwidth), and 32\,GB RAM. All timings are wall-clock measurements from the current codebase. Experiments presented here were run using release v1.1.0 of the software, archived on Zenodo (\url{https://zenodo.org/records/18837081}).

\subsection{CPU Baseline}

Table~\ref{tab:cpu} reports the performance of the sequential CPU oracle \texttt{cpu\_goldbach}. These results establish the reference against which GPU speedups are measured. The baseline scales linearly with~$N$.

\begin{table}[ht]
\centering
\caption{CPU baseline (\texttt{cpu\_goldbach}): sequential sieve and Goldbach scan. No failures at any limit.}
\label{tab:cpu}
\begin{tabular}{rrrrr}
\toprule
Limit & Even $n$ checked & Sieve & Check & Total \\
\midrule
$10^{7}$  & 4,999,999     & 27\,ms     & 89\,ms      & 117\,ms \\
$10^{8}$  & 49,999,999    & 768\,ms    & 1,052\,ms   & 1,820\,ms \\
$10^{9}$  & 499,999,999   & 9,836\,ms  & 12,786\,ms  & 22,622\,ms \\
$10^{10}$ & 4,999,999,999 & 119,947\,ms & 188,388\,ms & 308,335\,ms \\
\bottomrule
\end{tabular}
\end{table}

\subsection{Range Verification and Memory Efficiency}

Table~\ref{tab:range_gpu2} presents results for \texttt{goldbach\_gpu2}, which uses a global GPU-resident bitset. Relative to the CPU baseline, \texttt{goldbach\_gpu2} achieves a $16\times$ speedup at $10^{9}$ and a $12\times$ speedup at $10^{10}$. At smaller limits ($10^{7}$--$10^{8}$), kernel launch overhead reduces the GPU advantage, but from $10^{9}$ upward the GPU consistently dominates.

\begin{table}[ht]
\centering
\caption{Range verification using \texttt{goldbach\_gpu2} (global bitset, OpenMP sieve, 20 threads). No failures. Cannot reach $10^{11}$: bitset would require 5,960\,MB, leaving insufficient headroom on an 8\,GB device.}
\label{tab:range_gpu2}
\begin{tabular}{rrrrr}
\toprule
Limit & Even $n$ checked & Sieve (CPU) & Kernel (GPU) & Total \\
\midrule
$10^{7}$  & 4,999,999         & 16\,ms     & 6\,ms      & 22\,ms \\
$10^{8}$  & 49,999,999        & 46\,ms     & 75\,ms     & 121\,ms \\
$10^{9}$  & 499,999,999       & 687\,ms    & 698\,ms    & 1,386\,ms \\
$10^{10}$ & 4,999,999,999     & 17,923\,ms & 7,520\,ms  & 25,443\,ms \\
\bottomrule
\end{tabular}
\end{table}

The memory footprint of the dense bitset is summarised in Table~\ref{tab:memory}. The $16\times$ reduction relative to a byte array makes GPU verification feasible up to $10^{11}$, but even the bitset becomes infeasible at $10^{12}$, motivating the segmented design of \texttt{goldbach\_gpu3}.

\begin{table}[ht]
\centering
\caption{VRAM usage: byte array vs.\ dense bitset. The $16\times$ reduction is constant. At $10^{12}$, segmentation (\texttt{goldbach\_gpu3}) is required.}
\label{tab:memory}
\begin{tabular}{rrrl}
\toprule
Limit & Byte array & Bitset & Feasible on 8\,GB GPU? \\
\midrule
$10^{9}$  & 953\,MB     & 59\,MB     & Both \\
$10^{10}$ & 9,536\,MB   & 596\,MB    & Bitset only \\
$10^{11}$ & 95,367\,MB  & 5,960\,MB  & Bitset only \\
$10^{12}$ & 953,674\,MB & 59,605\,MB & Neither \\
\bottomrule
\end{tabular}
\end{table}

\subsection{Segmented Range Verification}

Table~\ref{tab:range_gpu3} reports results for \texttt{goldbach\_gpu3} with $P_{\text{SMALL}} = 10^6$ (78,498 GPU-resident primes). No failures were observed at any limit. All segments were resolved entirely in Phase~1, with zero CPU fallbacks, implying that $p_{\min}(n) \le 2\times 10^6$ for all even $n \le 10^{12}$. This is consistent with the envelope prediction $H(10^{12}) \approx 2000$ from Equation~\ref{eq:hx}, and our bound exceeds the predicted worst case by three orders of magnitude.

Although \texttt{goldbach\_gpu3} is slower than \texttt{goldbach\_gpu2} at shared limits due to per-segment sieve construction and VRAM transfers, it is the only design capable of reaching $10^{12}$ and beyond on an 8\,GB GPU.

\begin{table}[ht]
\centering
\caption{Exhaustive verification using \texttt{goldbach\_gpu3}
(segmented, \(P_{\text{SMALL}} = 10^6\), \\
\texttt{SEG\_SIZE} = \(10^7\)).
Phase~2 CPU fallbacks: zero at all limits. No failures.}
\label{tab:range_gpu3}
\begin{tabular}{rrrr}
\toprule
Limit & Even $n$ checked & Phase~1 success & Total time \\
\midrule
$10^{9}$  & 499,999,999     & 100\% & 2,048\,ms \\
$10^{10}$ & 4,999,999,999   & 100\% & 23,547\,ms \\
$10^{11}$ & 49,999,999,999  & 100\% & (3.8 minutes) 226,490\,ms \\
$10^{12}$ & 499,999,999,999 & 100\% & (41 minutes) 2,457,040\,ms \\
\bottomrule
\end{tabular}
\end{table}

\subsection{Runtime Scaling}

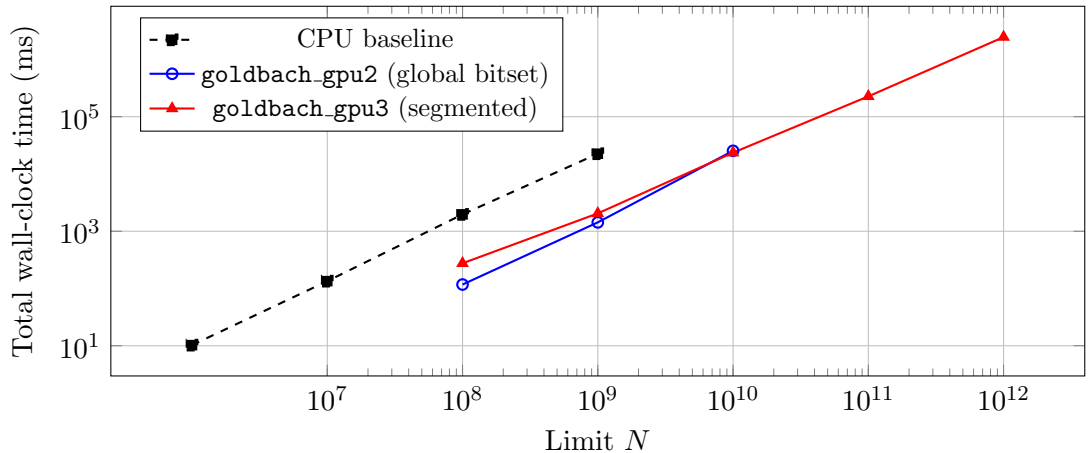
\begin{figure}[ht]
\centering
\begin{tikzpicture}
\begin{loglogaxis}[
    width=0.88\linewidth,
    height=6.5cm,
    xlabel={Limit $N$},
    ylabel={Total wall-clock time (ms)},
    legend pos=north west,
    legend style={font=\small},
    grid=major,
    xtick={1e7,1e8,1e9,1e10,1e11,1e12},
    xticklabels={$10^7$,$10^8$,$10^9$,$10^{10}$,$10^{11}$,$10^{12}$},
]
\addplot[mark=square*, color=black, dashed, thick] coordinates {
    (1000000, 10.2465)
    (10000000, 134.314)
    (100000000, 1967.71)
    (1000000000, 22568.8)
};
\addlegendentry{CPU baseline}

\addplot[mark=o, color=blue, thick] coordinates {
    (100000000, 117.223)
    (1000000000, 1427.47)
    (10000000000, 25358)
};
\addlegendentry{\texttt{goldbach\_gpu2} (global bitset)}

\addplot[mark=triangle*, color=red, thick] coordinates {
    (100000000, 275.34)
    (1000000000, 2048.58)
    (10000000000, 23546.7)
    (100000000000, 226490)
    (1e12, 2457040)
};
\addlegendentry{\texttt{goldbach\_gpu3} (segmented)}

\end{loglogaxis}
\end{tikzpicture}
\caption{Log--log runtime scaling. CPU baseline (black dashed), \texttt{goldbach\_gpu2} (blue, global bitset, limited to $10^{10}$ in practice), and \texttt{goldbach\_gpu3} (red, segmented, no VRAM ceiling). All three scale approximately linearly. \texttt{goldbach\_gpu3} is slower at shared limits due to per-segment overhead, but is the only tool capable of reaching $10^{12}$.}
\label{fig:scaling}
\end{figure}

Figure~\ref{fig:scaling} shows log--log runtime for all three tools. Both GPU implementations scale linearly with~$N$, consistent with the $O(N)$ behaviour predicted in Section~\ref{sec:gpu3}. The segmented design exhibits near-constant per-segment cost as $N$ increases from $10^{9}$ to $10^{12}$, confirming the theoretical analysis.

\subsection{Arbitrary Precision Results}

Table~\ref{tab:big} reports results for \texttt{big\_check}. For moderately large inputs, runtime is governed primarily by the index of the first valid prime $p$: a 204-digit number requiring $p = 113$ is verified in 43\,ms, while a 1,001-digit number requiring $p = 26{,}981$ takes 363\,ms. For very large inputs, however, the $O(d^3)$ cost of GMP's Miller--Rabin test becomes dominant, and digit count increasingly determines total runtime. The synchronised batch strategy reduces the $10^{10000}$ runtime from 231\,s (naïve OpenMP) to 182\,s, a 21\% improvement by preventing intra-batch thread runaway.

\begin{table}[ht]
\centering
\caption{Arbitrary precision verification (\texttt{big\_check}, GMP, 20 OpenMP threads, batch size 1,000). Timing is dominated by the index of the first valid \(p\) for moderate digit counts; for very large inputs, the \(O(d^3)\) cost of GMP primality tests becomes dominant.}
\label{tab:big}
\begin{tabular}{rrrr}
\toprule
$n$ & Digits & $p$ found & Time \\
\midrule
$10^{50}$    & 51     & 383    & 43\,ms \\
$10^{100}$   & 101    & 797    & 43\,ms \\
$10^{200}$   & 204    & 113    & 43\,ms \\
$10^{1000}$  & 1,001  & 26,981 & 363\,ms \\
$10^{10000}$ & 10,001 & 47,717 & 182,000\,ms \\
\bottomrule
\end{tabular}
\end{table}

\subsection{Multi-GPU Scaling and Efficiency}
The \texttt{goldbach\_gpu3} binary accepts a \texttt{--gpus=i} flag that distributes work across the requested GPUs (or \texttt{--gpus=-1} for all available devices). Strong-scaling experiments were performed on two platforms: a consumer-grade node with up to 6$\times$ NVIDIA RTX 3090 GPUs and a data-centre node with NVIDIA H100 GPUs.

At $N=10^{10}$ the framework exhibits near-linear scaling on the RTX 3090 system (Table~\ref{tab:scaling}). Efficiency remains above 91\% up to 4 GPUs and drops only modestly to 80.1\% at 6 GPUs, consistent with the onset of Amdahl’s Law as the CPU segmented-sieve stage begins to limit throughput.

On the H100 platform (single GPU baseline), wall-clock time at $N=10^{10}$ is essentially identical to a single RTX 3090 ($\approx$19.2\,s vs.\ 20.6\,s): still CPU-dominated. At $N=10^{11}$ with 4 H100 GPUs the run completes in 49.86\,s (versus 54.0\,s on 4$\times$ RTX 3090). These results indicate that larger verification limits ($N\gtrsim 10^{11}$) and architecture-specific tuning of \texttt{SEG\_SIZE} and \texttt{PSMALL} are required to leverage the superior memory bandwidth and compute of data-centre hardware.

\begin{table}[ht]
\centering
\caption{Strong Scaling at $N=10^{10}$ on NVIDIA RTX 3090 (\texttt{SEG\_SIZE} = $10^7$).}
\label{tab:scaling}
\begin{tabular}{cccc}
\toprule
GPUs & $N$ & Total Time (s) & Parallel Efficiency \\
\midrule
1   & $10^{10}$  & 20.57 & 100\% \\
2   & $10^{10}$  & 10.52 & 97.7\% \\
4   & $10^{10}$  & 5.64  & 91.2\% \\
6   & $10^{10}$  & 4.28  & 80.1\% \\
\midrule
4   & $10^{11}$  & 54.0  & -- \\
\bottomrule
\end{tabular}
\end{table}

\begin{table}[ht]
\centering
\caption{H100 Scaling Results ($N=10^{10}$ and $N=10^{11}$).}
\label{tab:h100-scaling}
\begin{tabular}{lcc}
\toprule
Configuration & $N$ & Total Time (s) \\
\midrule
1$\times$ H100   & $10^{10}$ & 19.18 \\
2$\times$ H100   & $10^{10}$ & 11.04 \\
4$\times$ H100   & $10^{10}$ & 5.42  \\
\midrule
4$\times$ H100   & $10^{11}$ & 49.86 \\
\bottomrule
\end{tabular}
\end{table}

\section{Discussion}
\label{sec:discussion}

\subsection{Architectural vs.\ performance contribution}

The results demonstrate that the VRAM limitations reported in prior GPU-based work~\cite{varon2020} arise from monolithic prime-table storage rather than from any inherent constraint of GPU architectures. The dense bitset representation removes this bottleneck up to $10^{11}$, and the segmented double-sieve design removes it entirely. Together, these techniques enable exhaustive verification up to $10^{12}$ on a single consumer GPU—an order of magnitude beyond previous single-GPU efforts—without distributed infrastructure or specialised hardware.

\subsection{Empirical validation of the minimal Goldbach prime envelope}

The 100\% Phase~1 success rate across all segments of the $10^{12}$ run provides an exhaustive empirical confirmation that $p_{\min}(n) \le 2\times 10^6$ for all even $n \le 10^{12}$. This is consistent with the upper-envelope prediction $H(10^{12}) \approx 2000$ from Equation~\ref{eq:hx}, and reinforces the Goldbach comet picture: the number of Goldbach partitions $c(n)$ grows with $n$, ensuring that small-prime partitions remain abundant throughout the tested range.

\subsection{Algorithmic gap with the state of the art}

The algorithm implemented in \texttt{goldbach\_gpu3}—iterating over candidate primes $p$ and testing whether $n-p$ is prime—is in the spirit of Oliveira~e~Silva's double-sieve. However, our approach still performs per-$n$ primality lookups and occasional Miller--Rabin tests. In contrast, the full double-sieve method applies bitwise bulk-marking across entire segments, eliminating per-$n$ primality tests and achieving $O(N \log \log N)$ complexity with extremely small constants.

Adapting this approach to GPUs would require representing the verified array as a bitset and replacing per-$n$ lookups with bitwise AND/OR operations across aligned words. Such a design would eliminate the Miller--Rabin fallback path and substantially reduce per-segment work. We identify this as the primary direction for future work and the most promising route toward reaching $10^{15}$ on a single GPU.

\subsection{Limitations}
The dominant runtime cost remains the CPU-based segmented sieve, which must rebuild the prime table for every segment before transferring it to the GPU(s). Even on an H100, a single-GPU run at $N=10^{10}$ takes essentially the same wall-clock time ($\approx$19.2\,s) as a single RTX 3090, because the GPU kernel finishes in a few hundred milliseconds and then idles while waiting for the next CPU-generated segment.

Performance is also sensitive to the choice of \texttt{SEG\_SIZE} and \texttt{PSMALL} relative to the host CPU’s cache and memory bandwidth. When the segment bitset fits in L2/L3 cache ($\approx$600\,KB at \texttt{SEG\_SIZE} $=10^7$), parallel efficiency exceeds 90\% on up to 4 GPUs. Larger segments that spill into DRAM expose the host’s $\approx$50--100\,GB/s memory bandwidth limit, starving even high-bandwidth GPUs (448\,GB/s on RTX 3090, 3.35\,TB/s on H100). Consequently, data-centre hardware only begins to show its advantage at larger verification limits ($N\gtrsim10^{11}$).

For \texttt{big\_check}, the $O(d^3)$ cost of GMP primality tests limits practical verification to approximately $10^{10,000}$ within a few minutes on 20 threads. At $10^{100,000}$, individual primality tests take seconds, making exhaustive batch search impractical.

\subsection{Future work}
Promising directions include GPU-accelerated segmented sieve construction (to eliminate the host-device dependency), an Oliveira-style bitwise bulk-marking kernel targeting $O(N \log \log N)$ complexity, large-scale computation of Goldbach partition counts $c(n)$, and architecture-specific tuning of \texttt{SEG\_SIZE}/\texttt{PSMALL} together with deployment on high-memory GPUs to push the exhaustive frontier.


\section{Conclusion}
\label{sec:conclusion}

This work presents an open-source framework that achieves exhaustive verification of Goldbach’s conjecture up to $10^{12}$ on a single consumer GPU in under 2 hours. A dense bitset representation reduces VRAM requirements by $16\times$, enabling GPU-resident prime tables and a $16\times$ speedup over the CPU baseline at $10^{9}$. The segmented double-sieve verifier \texttt{goldbach\_gpu3} removes the hard VRAM ceiling: each segment requires only $14$\,MB regardless of total range, and the design achieves $O(N)$ wall-clock time with $O(1)$ memory in $N$. A rigorous CPU fallback guarantees mathematical completeness; in all experiments it was never invoked. An arbitrary-precision GMP checker extends single-number verification to $10^{10000}$ in 182\,s on 20 threads. No counterexamples were found in any of the computations.


\section*{Software Availability and Installation}
\label{sec:data}

The framework targets Linux-based systems equipped with the NVIDIA CUDA Toolkit ($\geq 11.0$), the GMP library, and an OpenMP-capable C++ compiler. The build process is managed via CMake:

\begin{lstlisting}[language=bash]
mkdir build && cd build
cmake .. && make -j$(nproc)
./bin/goldbach_gpu3 1000000000000
\end{lstlisting}

The repository (\url{https://github.com/isaac-6/goldbach-gpu}) includes comprehensive documentation, a validation suite, and reproducible example runs to ensure consistent behaviour across different GPU architectures. The project is released under an open-source license as specified in the repository.

\section*{Acknowledgements}

The author thanks the developers of CUDA, GMP, and OpenMP for the foundational libraries that made this work possible. The author also acknowledges the seminal computational work of Oliveira~e~Silva, Herzog, and Pardi~\cite{oliveira2014}, which established the algorithmic context and baseline for this project.


\end{document}